 \documentclass[article]{elsart}

\usepackage{amssymb}

\begin{document}
\newbox\Ancha
\def\gros#1{{\setbox\Ancha=\hbox{$#1$}
   \kern-.025em\copy\Ancha\kern-\wd\Ancha
   \kern.05em\copy\Ancha\kern-\wd\Ancha
   \kern-.025em\raise.0433em\box\Ancha}}

\begin{frontmatter}

% Title, authors and addresses

% use the thanksref command within \title, \author or \address for footnotes;
% use the corauthref command within \author for corresponding author footnotes;
% use the ead command for the email address,
% and the form \ead[url] for the home page:
% \title{Title\thanksref{label1}}
% \thanks[label1]{}
% \author{Name\corauthref{cor1}\thanksref{label2}}
% \ead{email address}
% \ead[url]{home page}
% \thanks[label2]{}
% \corauth[cor1]{}
% \address{Address\thanksref{label3}}
% \thanks[label3]{}

\title{ An algebraic 
$SU(1,1)$ solution for the relativistic hydrogen atom}

% use optional labels to link authors explicitly to addresses:
% \author[label1,label2]{}
% \address[label1]{}
% \address[label2]{}

\author{R. P. Mart\'{\i}nez-y-Romero}

\address{Facultad de Ciencias, Universidad Nacional Aut\'onoma de M\'exico,  Aparta\-do Pos\-tal 21-267, C P 04000, Coyoac\'an D. F. M\'e\-xico.}
\ead{rmr@hp.fciencias.unam.mx}
\author{H. N. N\'u\~nez-Y\'epez}
\address{Departamento F\'{\i}sica,  Universidad
 Aut\'onoma Metropolitana, Unidad Izta\-pa\-lapa, Apar\-tado Postal 55-534, C P 09340,
Iztapalapa D. F. M\'exico.}
\ead{nyhn@xanum.uam.mx}
\author{A. L. Salas-Brito}
\address{Laboratorio de Sistemas Din\'amicos, Departamento de Ciencias
B\'a\-sicas, Universidad Aut\'onoma Metropolitana, Unidad Azca\-pot\-zal\-co,
Aparta\-do Pos\-tal 21-267, C P 04000, Coyoac\'an D. F. M\'e\-xico.}
\ead{asb@correo.azc.uam.mx}

\begin{abstract}
The bound eigenfunctions and  spectrum of a Dirac hydrogen atom are found  taking advantage of the $SU(1, 1)$ Lie algebra in which the radial part of the problem can be expressed.  For defining the algebra we need to add  to the description an additional angular variable playing essentially the role of a phase. The operators spanning the algebra are used for defining ladder operators for the radial eigenfunctions of the relativistic hydrogen atom and for evaluating its energy spectrum.  The status of the Johnson-Lippman operator  in this algebra is also investigated. 
\end{abstract}

\end{frontmatter}

The bound solutions of the hydrogen atom are of great importance in 
both classical  and  quantum mechanics and so is the search for new 
ways of solving or using 
such problem \cite{castano,jllb05,dong04,mci,ma03,draganescu,sous99,sous98,kerala04,gortel91}. 
 The purpose of this paper is to discuss an algebraic solution for the bounded eigenstates of the relativistic hydrogen atom  based on the $SU(1,1)$  properties of the extended radial part of the problem \cite{li00,wyb}. The  algebra is spanned by  two  operators $\Xi_i,\;i=1,2$ in terms of which such part of the problem can be expressed, plus an additional ---hence, {\sl extended}---  phase shifting operator, $\Xi_3$, whose only effect is to change the relative phases of the eigenfunctions of the algebra participating in  its radial  eigenfunctions; though,  for maintaining the  eigenfunctions of the hydrogen atom, it is found necessary to project our results to  vanishing additional phase after applying the operator $\Xi_3$.  That notwithstanding, the  set of operators allow  a very direct solution of the relativistic hydrogen problem. Moreover, the solution exhibits that the extended radial system possesses a $SU(1,1)$ symmetry\footnote{symmetry that can,  by suitably changing the definition of the $\Xi_i$---hence the conmutation relations---be recasted as a $SU(2)$ one\cite{sous99,li00,torres04}}---the original problem possessing an $SO(2,1)$ one \cite{alhaidari02}. We further exhibit how the eigenfunctions of the algebra, {\it i. e.} the radial eigenfunctions of  the relativistic hydrogen atom, can be related to the  Laguerre polynomials of noninteger indices and to the Sonine polynomials \cite{magnus}. These results can be of importance in the study of squeezed and coherent states of the relativistic hydrogen atom, as well as in diverse quantum optics applications \cite{draganescu,vourdas90,gour04,coherent99}. 

That the relativistic Coulomb problem could be solved exactly and that the energies so calculated showed an excellent  agreement with the experiments was indeed  one of the reasons for the rapid acceptance of Dirac relativistic quantum mechanics.   Almost from the start operational methods were applied to the hydrogen atom, as in \cite{temple30,sauter35,davis39}. In due course  other  methods were developed, for example in  \cite{hylleras55,kolsrud66}, and in Biedenharn's solution \cite{biedenharn62} in which advantage was taken of the constant operator $B$ introduced by Johnson and Lippmann  \cite{biedenharn62,jl,delange}---shown in equation (\ref{jl}) below. 

The  Hamiltonian of the relativistic hydrogen atom is

\begin{equation}\label{Eq.1}
H_D = c {\gros\alpha}\cdot\hbox{\bf p} + \beta c^2 - {1\over 
r},
\end{equation}

 using atomic units ($\hbar=m=e=1$), $c$ is  the speed of light, ${\gros\alpha}$ and $\beta$ are the  Dirac matrices in the Dirac representation \cite{bjorken,2},

\begin{equation}\label{equation2}
 {\gros \alpha}= \pmatrix{ 0&  {\gros\sigma}\cr
                           {\gros\sigma}&   0 }, \qquad \beta=\pmatrix{1& 0\cr
               0& -1}, 
\end{equation}

\noindent where the 1's and 0's stand, respectively, for $2\times2$ unit and zero matrices and the $\mathbf\sigma$ is the operator composed by the three Pauli matrices ${\gros \sigma}=(\sigma_x, \sigma_y, \sigma_z)$. The method to be discussed in this paper is  different from all the aforementioned, as it deals exclusively with the radial part of the problem. Besides, to obtain a system of equations invariant under $SU(1,1)$, we need to extend the configuration space by adding an extra angular variable $\varphi$ (Cf.\ equation (\ref{20}) below).   

Since the Hamiltonian (\ref{Eq.1}) is invariant under rotations, we look for simultaneous eigenfunctions of  $H_D$, $|{\bf J}|^2$ and $J_z$, where ${\bf J}= {\bf L} + {\bf s}$ is the total angular momentum,  and  ${\mathbf s}\equiv {1\over 2}{\bf \Sigma}$ is the spin. Parity is a good quantum number in our problem, but  instead of  parity, we  use  the parity related quantum number $\epsilon$, defined as
\begin{equation}\label{5} 
\epsilon =\cases{1 & If $ l=j + {1\over 2},$\cr
\cr
-1 & If $ l= j- {1\over 2}$,}
\end{equation}

\noindent we can easily also show that the eigenvalues of the operator $K\equiv\beta(1 + {\gros\Sigma}\cdot{\bf L}),$ are given by  $-\epsilon(j + 1/2),$ so, in a way,  $\epsilon$ is a more useful quantity  than parity.

The lower components of the wave function should have  an opposite parity than the upper ones, and as parity goes as $(-1)^l$, where $l$ is the orbital angular momentum quantum number,  if we  define $l'=j-\epsilon/2$,  we can cast the eigensolutions of (\ref{Eq.1})  as  

\begin{equation}\label{6} 
\Psi(r,\theta,\phi) = {1\over r}\left( \matrix{F(r){ \mathcal   Y}_{j=l-\epsilon/2}^m(\theta, \phi)\cr \cr iG(r){\mathcal    Y}_{j=l' + \epsilon/2}^m(\theta,\phi)}\right). 
\end{equation}
 
\noindent where $ {\mathcal  Y}_{j=l\pm\epsilon/2}^m(\theta, \phi)$ are spinorial harmonics \cite{greiner}. The form proposed for the eigenfunctions solves  the angular part of the hydrogen atom. 
 
 Let us  define the quantity 
$k\equiv \sqrt{ c^4 - E^2}$, which is positive definite since we are  interested in bound states. Furthermore, let us also define 

\begin{equation}\label{7} 
 \zeta\equiv \alpha_F=1/c, \quad \tau_j\equiv \epsilon\left(j+{1\over 2}\right), \quad \nu \equiv \sqrt{c^2-E\over c^2+ E},
\end{equation}
 
\noindent where $\alpha_F \simeq 1/137$ is the fine structure constant.  With the proposed solution (\ref{6}), we have to deal only with the radial equation for the relativistic hydrogen atom \cite{ajp00}

\begin{equation}\label{radialha} 
 {H_D}_r = \alpha_r \, 
\left [ p_r -i{\tau_j\over r}\beta \right] 
+ c^2 \beta - \frac{1} {r},
\end{equation}

where

\begin{equation}\label{defs}
\alpha_r \equiv {1\over r} {\gros\alpha}\cdot{\bf r},\qquad
  p_r \equiv  - { i \over r}\left(1 + r{d\over dr}\right).
\end{equation}

Using the radial part of the solution (\ref{6}),  definitions (\ref{7}) and (\ref{defs}), equation (\ref{radialha}), and using $\rho\equiv kr$, we can obtain the pair of equations
\cite{sous98}

\begin{equation}\label{14a}
\left(-{d\over d\rho} + {\tau_j\over \rho}\right)G(\rho) = \left( -\nu 
+{\zeta\over \rho}\right)F(\rho),
 \end{equation}

\noindent and

\begin{equation}\label{14b}
 \left(+{d\over d\rho} + 
{\tau_j\over\rho}\right)F(\rho) = \left( \nu^{-1} + 
{\zeta\over\rho}\right)G(\rho).
 \end{equation}

 The above equations, (\ref{14a}) and (\ref{14b}), can be rewritten using a set of three operators whose commutation relations define a $SU(1,1)$ algebra, as we exhibit in what follows. Introducing the  variable $x$ through the relation $ \rho=e^x$, so $x\in (-\infty, \infty)$, and  the functions $\psi_{\pm}(x)$ through

\begin{eqnarray}\label{16a}
F(\rho(x))  = \sqrt{c^2 + E}\,\left[\psi_{-}(x) + \psi_{+}(x)\right], \\
 G(\rho(x))  = \sqrt{c^2 - E}\,\left[\psi_{-}(x) - \psi_{+}(x) \right], 
\end{eqnarray}

\noindent we  obtain 

\begin{equation}\label{17a}
\left[{d\over dx} + e^x - {\zeta E\over\sqrt{c^4 - E^2}} 
\right]\psi_+(x)  =  \left({\zeta \over \sqrt{c^4 - E^2}} - \tau_j
\right)\psi_-(x),
\end{equation}

\noindent and 

\begin{equation}\label{17b} 
-\left[{d\over dx} -e^x + {\zeta 
E\over\sqrt{c^4 - E^2}} \right]\psi_-(x)  =  \left({\zeta \over 
\sqrt{c^4 - E^2}} + \tau_j\right)\psi_+(x).
 \end{equation}

\noindent This first-order  system can be uncoupled   to yield  

\begin{equation}\label{18a} 
\left[{d^2\over dx^2} +  2\mu e^x - e^{2x}-
{1\over 4} \right]\psi_+(x)  =  \left( {\tau_j}^2 -\zeta^2 - {1\over 
4}\right)\psi_+(x),
 \end{equation}

\noindent and

\begin{equation}\label{18b}
\left[{d^2\over dx^2}+
2\left(\mu -1\right)e^x  - e^{2x} -{1\over 4} \right]\psi_-(x)  = 
\left(\tau_j^2 -\zeta^2 - {1\over 4}\right)\psi_-(x), 
\end{equation}

\noindent where we have defined 

\begin{equation}\label{E}
\mu \equiv {\zeta E\over \sqrt{c^4 - E^2 }} + 1/2.
\end{equation}

 The seemingly odd $-1/4$ term  in (\ref{18a}) and (\ref{18b}), has to be included in order to close the algebra that solves the problem. We pinpoint that the change of variable from $\rho$ to $x$ is, strictly speaking, not necessary for any of the calculations that follow;  the change, however, simplifies the appearance of some of the equations.  Notice also that we can regard the equations (\ref{18a}) and (\ref{18b}) as  eigenvalue equations with a  common and predetermined eigenvalue $\xi =\tau_j^2 -\zeta^2 - {1\over 4} = j(j+ 1) - \zeta^2$.  Since the minimum value of $j$ is  $1/2$, then $\xi > 0$. In cases where the atomic number, $Z$, is different from 1 ---the value we are assuming in this article--- then $\xi\ge 0$ for  $Z= 2,\cdots $ up to 118. If $Z$ exceeds this value then $\xi$ could become imaginary and the solutions would then  exhibit  irregular  behaviour \cite{delange}. 

To construct the Lie algebra for our system, let us  introduce an extra  variable, $\varphi \in [0,2\pi]$,  through the operator

\begin{equation}\label{20}
\Xi_3 \equiv -i{\partial \over \partial \varphi},
 \end{equation}

\noindent  and then define 

\begin{equation}\label{21}
\Xi_{\pm} \equiv ie^{\pm i\varphi}\left({\partial \over \partial x} \mp 
e^x \mp i{\partial\over \partial \varphi} + {1\over 2}\right), 
\end{equation}

\noindent which depend both on $\varphi$ and on the transformed `radial' variable $x.$ These  operators satisfy the algebra 

\begin{equation}\label{22a}
\left[\Xi_3,\Xi_{\pm}\right] = \pm \Xi_{\pm}, \quad  \left[ \Xi_+, \Xi_- \right]=  -2\Xi_3. 
\end{equation}

Let us now introduce  the operators $\Xi_1$ and $\Xi_2$ 

\begin{equation}\label{23}
\Xi_1= {1\over 2} \left(\Xi_+ + \Xi_-\right), \qquad \Xi_2= {1\over 2i} \left(\Xi_+ - \Xi_-\right), 
\end{equation}

\noindent which together with   $\Xi_3$ satisfy the  $SU(1,1) $  Lie algebra \cite{mci,wyb},

\begin{equation}\label{24}
[ {\Xi}_1, {\Xi}_2]=-i {\Xi}_3,\quad  [ {\Xi}_2, {\Xi}_3]=i {\Xi}_1,\quad  [ {\Xi}_3, {\Xi}_1]=i {\Xi}_2;
\end{equation}

\noindent whose Casimir operator  is \cite{wyb} 

\begin{equation}\label{casimir}
\Xi_c = -\Xi^2_1 - \Xi_2^2 +\Xi_3^2 = {\partial^2/ \partial x^2} - e^{2x} -2i e^x 
{\partial/\partial \varphi}- {1/4}.
\end{equation}

 Note that we do not have  a linear term in $\partial/\partial x$ in the Casimir operator precisely because we included the $1/2$--term in the definition of the $\Xi_{\pm}$. 
It is also worth noting that the realization of the $SU(1,1)$ algebra given in the previous equations is in terms of two variables (another example is described in [31]) and not in terms of a single one as it is more usual \cite{bargmann,basu}.

To find the simultaneous eigenfunctions and eigenvalues of the operators $\Xi_c$ and $\Xi_3$,  let us  change to the variable $x,$ using $\rho =e^x$, and   write  the eigenfunctions as $V_\xi^\mu(x,\varphi)$; where $\mu$ and $\xi$ stand for the eigenvalues of $\Xi_3$ and $\Xi_c$, respectively, 

\begin{eqnarray}
\Xi_3 V_\xi^\mu (x,\varphi) &= \mu V_\xi^\mu (x,\varphi) \cr \cr
\Xi_c V_\xi^\mu (x,\varphi) &= \xi V_\xi^\mu(x,\varphi).
\end{eqnarray}

Using equation\ (\ref{20}), we  find 

\begin{equation}\label{29}
V_\xi^\mu(x,\varphi) = e^{i\mu \varphi}{    P}_\xi^\mu (x). 
\end{equation}

\noindent It is clear  that $\varphi$ plays the role of a phase, and 

\begin{equation}\label{30}
\psi_+(x) ={    P}_\xi^\mu(x), \qquad \psi_-(x) ={    P}_\xi^{\mu -1}(x).
 \end{equation}

 From equation (\ref{22a}), we can also conclude that the $\Xi_{\pm}$ are ladder operators for the problem,  they change  $\mu$ to  $\mu\pm 1$, so they are easily shown to satisfy 

\begin{equation}\label{31}
\Xi_{\pm}V_\xi^\mu (x,\varphi) \propto V_\xi^{\mu \pm 1}(x,\varphi). 
\end{equation}

To establish  the properties of the $\Xi$-operators  we need  to define the inner product 

\begin{equation}\label{32}
<\phi,\psi> = \int_0^{2\pi}{d\varphi\over 2\pi}\int_{-\infty}^\infty  
\phi^{*}(\varphi,x)\psi(\varphi,x)\,dx, 
\end{equation}

 where we are assuming that $  \phi(\varphi,x)$ and $ \psi(\varphi,x)$ are periodic functions over the interval $ \varphi\in[0,2\pi] $ and that they vanish as $x\to\pm\infty$---so as to make them square-integrable. Using (\ref{32}), we can prove that $\Xi_1, \Xi_2$, and $\Xi_3$ are hermitian, $\Xi_i^{\dag} =\Xi_i,\,  i=1,2,3$.  

To see  that the problem of finding  the radial eigenfunctions of the extended radial part of the relativistic hydrogen atom is  equivalent to finding a representation of the $SU(1,1)$ algebra, we   introduce  a complete orthogonal basis of simultaneous eigenfunctions for $\Xi_c$ and $\Xi_3$, namely  $V_\xi^\mu (x,\varphi)\equiv |\xi\,\mu>$,  we  require that $ <\xi'\,\mu' | \xi\,\mu> = \delta_{\xi ,\xi'}\,\delta_{\mu,\mu'}$,  using the inner product (\ref{32}). 

The operator  ${\mathbf\Xi}^2$  $ = \Xi_1^2 + \Xi_2^2 +\Xi_3^2$ is obviously definite positive; furthermore, as

\begin{equation}\label{39}
{\mathbf\Xi}^2=2\Xi_3^2-\Xi_c, 
\end{equation}

we must have $ 2\mu^2\ge \xi$, this in turn means that $|\mu|$ is bounded by below, as a consequence there is a minimum value for it, say $|\mu|_{min}$. We have then two choices, either we choose $\mu>0$ or we choose $\mu<0.$ In the first case  $\mu$ itself is bounded by below; in the second, $\mu$ is bounded by above. 

Let us take first $\mu>0$ and   define $\lambda\equiv \mu_{min}$. Let us call $\lambda\equiv|\mu|_{min}$  then $\Xi_-|\xi\,\lambda> =0$,  or equivalently $\Xi_+\Xi_-|\xi\,\lambda> =0$. Since $\Xi_+\Xi_- = -\Xi_c + \Xi_3^2- \Xi_3 $, this means 

\begin{equation}\label{41b}
-\xi +\lambda^2-\lambda=0 \quad \hbox{ or } \quad \xi= \lambda(\lambda-1). 
\end{equation}

As $\lambda$ has to be positive,  we directly obtain that $\mu_{\hbox{min}}\equiv \lambda=s+{1\over 2}$, where we have defined $s\equiv +\sqrt{\tau_j^2-\zeta^2}=+\sqrt{\left(j+{1/2}\right)^2-\alpha_F^2}$, so $s$ is real. 

 Some features of this algebraic approach are worth noting. Fist, notice that the eigenvalue $\xi= \lambda(\lambda -1) $ is not determined by the algebra itself, but just by  the angular symmetry of the problem.   Second, that this  eigenvalue is not necessarily an integer or half-integer number as it should be in the related $SU(2)$ case. Nevertheless, it is interesting to point out that in the non-relativistic limit $\lambda$ becomes approximately half-integer since in such a limit  we can neglect the fine structure constant $\alpha_F$,  so $s\approx (j+1/2)$. In this case the eigenvalues of $\Xi_c$ become $l(l+1)$ but the structure of the algebra $SU(1,1)$ remains unchanged. The remarkable point here is that the representation of the algebra and the energy spectrum become precisely those of the non-relativistic hydrogen atom \cite{mns}.

With the value of $\lambda$ fixed, we proceed to obtain the  representations of $SU(1,1)$.  We found it convenient to relabel the kets $ |\xi\,\mu> $, using $\xi =\lambda(\lambda -1)$, as $  |\lambda\,\mu> $. So, starting with $ |\lambda\,\lambda>$,  we obtain an infinite ---since $\mu$ is bounded only by below--- series of states  $  |\lambda\,\lambda+1>$,   $|\lambda\,\lambda+2>,\cdots,$ by aplying  $1, 2, \cdots$ times $\Xi_+$ to the ground state $  |\lambda\,\lambda>$. Therefore, $\mu$ itself takes the values $ \mu = \lambda, \,\, \lambda +1,\,\, \lambda +2,\cdots = \lambda + n,$ with  $ n=1,2, \cdots $ 

To find the energy spectrum,  we just need to solve, from definition (\ref{E}),  $E$ in terms of $\mu$;  thus  the energy spectrum of the hydrogen atom is 
 
\begin{equation}\label{43b} 
E=c^2\left[1 + {\zeta^2 \over (\mu -1/2)^2}\right]^{-1/2} =c^2\left[1 + { \alpha_F^2\over (\mu -1/2)^2}\right]^{-1/2},
\end{equation}

where  $\mu = \lambda +n$, or $\mu-1/2 = s +n$, $ s =  \sqrt{(j+1/2)^2 -\alpha_F^2}$,  $n= 0,1, 2,\cdots$.

Let us write the above reult in a more familiar form. We define the principal quantum number $N= 1,2,3 \cdots$ and the auxiliary quantity $\epsilon_j$ as follows

\begin{eqnarray} 
  & N \equiv j + 1/2 +n ,  \cr 
  \epsilon_j &\equiv N-s-n = j+1/2 -s.
\end{eqnarray}

From this definition we have that $\mu -1/2 = s +n = N -\epsilon_j.$ When we introduce this last result in the energy equation (\ref{43b}), we obtain precisely the Dirac energy spectrum for the Hydrogen atom,\cite{davis39,bjorken,2,ajp00,5}.

To obtain the eigenfunctions from the representation, let us  write
 
\begin{equation} 
\Xi_{\pm} |\lambda\,\mu>= C_\mu^{\pm}|\lambda\,\mu\pm 1>.
\end{equation}

We can evaluate these constants  from $<\lambda\,\mu|\Xi_+\Xi_-|\lambda\,\mu>=-\lambda(\lambda -1) + \mu(\mu -1)=C_{\mu-1}^{+}C_\mu^{-}$ and from $(C_{\mu-1}^{+})^*=C_\mu^{-}$; if we further assume these constants to be real, we get  $C_\mu^{\pm}=\pm\sqrt{\mu(\mu\pm1)-\lambda(\lambda-1)}$. 

The ground state wave function  is obtained from $\Xi_-|\lambda\,\lambda>=0$, that is  

\begin{equation}\label{50a}
{    P}^{\lambda}_\lambda(x)=N_\lambda\, e^{sx}\exp(-e^x)=N_\lambda \,e^{(\lambda-1/2)x}\exp(-e^x)= N_\lambda \rho^s e^{-\rho}, 
\end{equation}

\noindent where $ N_\lambda\equiv{ 2^{(\lambda-1/2)}\over \sqrt{\Gamma(2\lambda-1)} }$ is the normalization constant calculated using the inner product (\ref {32}) and $\Gamma(2\lambda-1)$ stands for the Euler-gamma function.  From equation (\ref{30}), we find that for the lowest eigenvalue, $ \psi_+(x)={    P}^{\lambda}_\lambda(x), $ and $ \psi_-(x) = 0$.
In terms of the radial variable $\rho$,  the ground state solutions  are [equation (\ref{16a})]

\begin{equation}\label{50c}
F(\rho)=N\sqrt{(c^2+E)\over 2c^2(\lambda-1/2)} \, \rho^s e^{-\rho},\;
G(\rho)=-N \sqrt{(c^2-E)\over 2c^2(\lambda-1/2)} \, \rho^s e^{-\rho}. 
\end{equation}

\noindent   The normalization constant for this state, $N= 1/\sqrt{2c^2(\lambda -1/2)}$, different from $N_\lambda,$  is obtained from the condition

\begin{equation}\label{51}
 \int_0^{\infty} \left( |F(\rho)|^2 + |G(\rho)|^2 \right) \, d\rho =1. 
\end{equation}

 The rest of the eigenfunctions for the positive set of eigenvalues $\mu$ are obtained by applying successively $ \Xi_+$  to the ground state $ |\lambda\,\lambda>$. Notice that the value of $\lambda$ is determined by the total angular momentum and is different for each value of $j$. 
 
The representation presented here is called the two-mode bosonic representation, because the operators (\ref{21}) can be constructed using two bosonic creation operators \cite{puri}.   This can be easily seen if we define the two mode bosonic operator operator $\hat\lambda \equiv 1/2(\hat a^\dagger \hat a - \hat b^\dagger \hat b +1),$  where $\hat a^\dagger$ and $\hat b^\dagger$ are the creation operators of (some) bosonic modes and $\hat a$ and $\hat b$ are the anhilation ones. If we put $ |m\,,n>$ as the simultaneous eigenstates of the operators $ \hat a^\dagger \hat a$ and $\hat b^\dagger \hat b ,$ respectively, we can see that our representation is given by $ |\lambda\,\mu> = |n + 2\lambda -1\,,n>,$ with $\mu =  \lambda +n, \, n= 0, 1,2\cdots $. In the non relativistic case $\lambda$ becomes an integer (the principal quantum number becomes an integer), because  $\lambda = l +1,$ \cite{mns} and the representation becomes what Bargmann calls {\it minimal M}, ($\lambda$ has a minimal value), $D_\lambda^+,$ where $\lambda $ takes the values $1/2,\, 1,\, 3/2,\, \cdots$\cite{bargmann}. In the non-relativistic case the semi-integer representations are excluded \cite{mns}. It is interesting that in the relativistic case we use a generalization to a non-integer, nor semi-integer index, of the minimal representation. 

What happens when we consider the negative set of eigenvalues of $\Xi_3$ ($\mu<0$)? In that case $ |\mu|_{min} =\lambda$ is the largest eigenvalue, instead of the smallest one. Consequently, we obtain the eigenfunctions by successive applications of $ \Xi_- $ to the ket $ |\lambda\,\lambda>$. Starting with this state and repeating the procedure we follow in the $\mu>0$ case we find that

\begin{equation}
 |\lambda\,\lambda> \sim e^{-i\lambda\xi}e^{sx}e^{e^x}= e^{-i\lambda\xi}\rho^se^{\rho}
\end{equation}

Which diverges as  $\rho\to \infty$. This behavior makes the negative eigenvalues solutions not square integrable, and we must have to discard them if we want to describe  physically realizable states, keeping only the positive set of eigenvalues of $\mu.$ .

As a last point, we note that in \cite{sous99} it is shown that all the solutions are polynomials with  weight factor $W(\rho)=\rho^{\lambda-1/2}e^{-\rho}$. This weight factor assures that the behavior of the big and the small components of the spinor are regular both at the origin as well as $\rho\to\infty$. 

We have thus constructed  an $SU(1,1)$ algebra for the relativistic  hydrogen atom by introducing the Hermitian operators  $\Xi_i$.   The eigenstates of the problem are  labelled by numbers $\mu,$ which are neither integers nor half-integers. Let us remark that we were forced to introduce the new variable $\varphi$ in order to close the algebra. In terms of the solutions of the Dirac equation, $\varphi$ just plays the role of an extra phase.   The functions, $\psi_\pm$, defining the eigensolutions of the hydrogen atom [Eqs.\ (\ref{16a}) and (11)] are, in strict, obtained by the projection to zero additional phase  of the eigenfunctions of the algebra, namely

\begin{equation}\label{53}
 \psi_+(x)={    P}_\xi^\mu(x)=  V_\xi^\mu (x,\varphi)|_{\varphi=0}, \;
 \psi_-(x)={    P}_\xi^{\mu -1}(x)=  V_\xi^{\mu-1} (x,\varphi)|_{\varphi=0}.
\end{equation}

What happens to the radial solutions $F(\rho)$ and $G(\rho)$ when we apply the operators $\Xi_c$ and $\Xi_3$ to the eigensolutions $  V_\xi^\mu (x,\varphi)$?  Given that both $  V_\xi^\mu (x,\varphi)$  and $ V_\xi^{\mu -1} (x,\varphi) $ are eigenfunctions of $\Xi_c$, the energy spectrum and eigenfunctions remain essentially unchanged, excepting for an irrelevant global constant $\lambda(\lambda-1)$. 
But, when we perform a $\Xi_3$ rotation things are different;  in this case we change $F(x)$ to

\begin{equation}\label{54}
F(x)\longrightarrow \hskip -14pt {}_{{}_{\Xi_3}}\,\, [e^{i(\mu-1)\varphi}\psi_-(x) + e^{i\mu\varphi}\psi_+(x)]= e^{i\mu\varphi}[e^{-i\varphi }\psi_-(x) + \psi_+(x)],
\end{equation}

\noindent since  $F(x)\sim [\psi_-(x) + \psi_+(x)]$. However, the energy spectrum does not change since both $ V_\lambda^\mu (x,\varphi)$  and $ \Xi_3V_\lambda^\mu (x,\varphi) $ have the same eigenvalue $\mu$.  Nevertheless,  as we can see from equation (\ref{54}), though the phase $e^{i\mu\varphi}$ does not play any observable role,   the term $e^{-i\varphi}$ changes the relative phase between the $\psi_+(x)$ and the $\psi_-(x)$  and so it changes the radial function $F(\rho)$ itself. This means that the rotated eigenfunction would no longer be an eigenstate of the radial hydrogen atom Hamiltonian unless we first project  the solution to vanishing phase: $\varphi\to 0$. We could say that there should be a sort of superselection rule between the functions $\psi_-(x)$ and  $\psi_+(x)$, this interpretation, apart from providing a way out of that minor problem,  would also confirm that the operator $\Xi_3$  cannot be an observable\cite{wick,supers}. All these  peculiar features of our solution method might have some bearings for the  properties of certain other systems \cite{today04}.

We  remark  that the eigenfunctions $\psi_+(x)= {    P}_\lambda^{\mu }(x)$   and $\psi_-(x)={    P}_\lambda^{\mu -1}(x)  $ are related to the Laguerre polynomials of non integer index \cite{davis39}, ${    L}_n^{2s}(2e^x)  $  and $ {    L}_{n-1}^{2s}(2e^x)$, as \cite{ajp00} 

\begin{equation}\label{55}
{    P}_\lambda^{\mu }(x)= a e^{sx}e^{-e^x}{    L}_n^{2s}(2e^x), \quad
 {    P}_\lambda^{\mu-1 }(x)=b e^{sx}e^{-e^x}{    L}_{n-1}^{2s}(2e^x) 
\end{equation}

\noindent where $\mu = n + s +1/2$, and the constants $a$ and $b$ should satisfy 

\begin{eqnarray}\label{56}
 a(\tau_j + s - \zeta\nu^{-1} + n) + b(n + 2s) =0,\;\cr
b( \tau_j - s + \zeta\nu^{-1} -n) -an =0.
\end{eqnarray}

 In fact, if we solve for $a$ in the second equation and substitute it back in the first of (\ref{56}),  using 
also that $s+n=\mu -1/2$,  we again obtain  precisely the energy spectrum of equation\ (\ref{43b}).
Notice that these  non-integer Laguerre polynomials are related to the hypergeometric function ${}_1F_1(-n,\alpha+ 1; x)$ and to the Sonine polynomials   $ T_\alpha^{\,(n)}(\rho)$  through the relations \cite{magnus}

\begin{eqnarray}\label{57}
 {    L}_n^{\alpha}(\rho)={\Gamma(\alpha+n+1)\over n! \Gamma(\alpha + 1)} {}_1F_1(-n,\alpha+1;\rho)\cr 
= (-1)^n{\Gamma(\alpha + n + 1) }\,T_\alpha^{\,(n)}(\rho).
\end{eqnarray}

Let us close the paper with some considerations about the  Johnson and Lippmann operator, $B$, defined as 

\begin{equation}\label{jl}
B={i\over c^2} K   \gamma_5(H_D -\beta c^2) -{1\over c}{ \Sigma}\cdot {\bf \hat r},
\end{equation} 
 
\noindent where $\gamma_5=i\alpha_1\alpha_2\alpha_3,$ and $K=\beta(1 + \Sigma\cdot{\bf L})$. This operator was introduced many years ago  to study the properties of  the Dirac Coulomb Hamiltonian \cite{jl,biedenharn62}.  A direct calculation shows that the  Hamiltonian of the relativistic hydrogen atom commutes with the Johnson and Lippmann operator, which can be regarded as the relativistic counterpart of the Laplace-Runge-Lenz vector of classical mechanics \cite{ijmp97}. 

To calculate the action of $B$ over the state functions (\ref{6}), we need to use the identities $ ({\gros\sigma}\cdot { \hat{\mathbf r}}){\mathcal Y}_{j=l\pm\epsilon/2}^m(\theta, \phi)= - {\mathcal Y}_{j=l\mp\epsilon/2}^m(\theta, \phi)$; after some algebra, we found that
 
 \begin{equation}
 B\Psi(r,\theta,\phi) = \pmatrix{\left[{\tau_j\over r}\left({dF\over dr} + {\tau_j F\over r} -{\zeta G\over r}\right) +{\zeta F\over r}\right]{\mathcal Y}_{j=l' + \epsilon/2}^m(\theta,\phi)\cr \cr
\left[{-i\tau_j\over r}\left({-dG\over dr} + {\tau_j G\over r} -{\zeta F\over r}\right) +{\zeta iG\over r}\right]{\mathcal Y}_{j=l + \epsilon/2}^m(\theta,\phi)}.
 \end{equation}

Here we see that the efect of this operator over the angular part interchanges the coupling of $l$ with $l'$ but the total angular momentum quantum number $j$ remains unchanged.  This point explains one of the degeneracies of the wave functions in the relativistic Coulomb problem, since the energy of a particular state depends on $N$ and $j$ and not on the coupling of the orbital angular momentum with  spin.  

We calculate now  the effect  of $B$ on the wave function  using the equations of motion (\ref{14a}) and (\ref{14b}), changing  to the variable $\rho = kr$ and writing only the effect over the radial part of the function

\begin{equation}\label{br}
 B \left[\frac {1} { \rho}\pmatrix{F(\rho)\cr iG(\rho)}\right] = \frac {1}{c^2 \rho} \pmatrix{c & - {i} \tau_j(c^2+E)\cr
 {i} \tau_j(c^2-E) & c } \pmatrix{F(\rho)\cr iG(\rho)}.
\end{equation}

\noindent As expected, the Johnson and Lippmann operator on shell is no longer a differential operator. Unfortunately, the operator is not diagonal in terms of $F$ and $G$; nevertheless, using  the functions $\psi_{\pm}$, we can write the above result as

\begin{equation}\label{psi}
 B \left[\frac {1} { \rho}\pmatrix{F(\rho)\cr iG(\rho)}\right] = \frac {1}{ \rho} \pmatrix{\sqrt{c^2 + E}
\left[(\zeta -\frac{k\tau_j}{c^2})\psi_+ + (\zeta + \frac{k\tau_j}{c^2})\psi_- \right]\cr
 \sqrt{c^2 - E}
\left[-(\zeta -\frac{k\tau_j}{c^2})\psi_+ + (\zeta + \frac{k\tau_j}{c^2})\psi_- \right]};
\end{equation}

 again we found a non diagonal result, nevertheless, from equation (\ref{30}), we know  that at least the ground state $ |\lambda\,\lambda>$ correspond to a diagonal case, because for that state $\psi_- =0$. Are there any other state in which this feature occurs? In order to obtain a diagonal state, we need that one of the two choices $ (\zeta \pm \frac{k\tau_j}{c^2})$ be null.
From  $\mu = s + 1/2 + n,\quad n=1,2,\cdots,$ $s=\sqrt{\tau_j^2 -\zeta^2}$ and equations (\ref{E}) and (\ref{43b}), we found that 

\begin{equation}
\left(\zeta \pm \frac{k\tau_j}{c^2}\right)= \zeta\left[1\pm \frac{\tau_j}{\sqrt{(s +n)^2 +\zeta^2}}   \right],
\end{equation}

which is never null excepting again for the ground state $ |\lambda\,\lambda>$. But this is {\it the ground state  of the representation of the $SU(1,1)$ algebra} and not of the hydrogen atom. In terms of the wave functions of the relativistic Coulomb Hamiltonian, the ground state of the algebra corresponds to the infinite set of states. Namely,  ($N=1$ and $j=0$), ($N=2$ and $j=1$), ($N=3$ and $j=2$) and so on. In other words, using  spectroscopy  notation $N_l$, we find the interesting result that the Johnson and Lippmann operator takes a diagonal form when written in terms of $\psi_\pm,$ only for the states 1s, 2p, 3d \dots

In conclusion, we have solved exactly the Dirac relativistic hydrogen atom using an algebraic technique based on the $SU(1,1)$ Lie algebra in terms of which the radial part of the problem can be expressed. With its help we calculated the explicit form of its ground state wave function. We have also derived radial ladder operators for the system and pointed out the way for obtaining all of its radial eigenfunctions using them. Let us also pinpoint that in the limit $\zeta \to 0$ the  algebra (\ref{22a}) [or (\ref{24})] trivializes, $\mu \to 1$, and the spectrum becomes continuous, as corresponds to a free Dirac particle. 

The method discussed in this paper, apart from the algebraic solution of the relativistic hydrogen atom, could be of interest in itself as it requires extending the system before the solution and then projecting the results back to  zero phase $\varphi=0$---although this last step is not always necessary. It  could be an interesting problem to account for all relativistic problems that may be solved by a similar extension and to find their corresponding extended symmetry algebras, as well as to investigate their generalization to $D>3$ spatial dimensions \cite{dong03}. The algebraic method  of solution and the expression of the bound state eigenfunctions in terms of the ladder operators $\Xi_\pm$   presented here may offer some  simplifications in   relativistic atomic physics calculations \cite{lombardi83}, can be useful in diverse applications in quantum optics \cite{draganescu,gour04} and it may, possibly,  be applied to the study of ionisation states \cite{bauer02}  by  using non-discrete representations of $SU(1,1)$; at least below the pair-production treshold, or in the non-relativistic limit  
\cite{howard91,smi85,gayet04}, where a description of the hydrogen atom in terms of such an $SU(1,1)$ algebra is also possible \cite{mns}.  

As we already said before, other application  of the method described here concerns the non-relativistic problem \cite{howard91,smi85,gayet04}, where a description of the hydrogen atom in terms of such an algebra is also possible.  The point here is that the $SU(1,1)$ algebra remains exactly the same also in the non--relativistic limit, and only the relationship of $\mu$ with the energy, the value of $s$ itself, and the fact that we have to deal with only one radial function  instead of two,  are the only important points we need to change \cite{mns}. 

\vskip 6 pt

 This work has been partially supported by a PAPIIT-UNAM grant (108302). We  acknowledge with thanks the comments and suggestions of   G. Schr\"odie, L. S. Micha, P. A. Koshka, G. Billi, C. Suri, G. R. Inti, M. Crida, D. Gorbe, M. Dochi, S. Mikei, G. R. Maya, L. Bidsi, M. Minina, and F. C. Sadi. ALSB also thanks the Centro Internacional de Ciencias (Cuernavaca) for the interesting environment in which this paper was revised.

\end{document}